\documentclass[a4paper,11pt]{article}
\usepackage{graphicx} 

\usepackage[english]{babel} 

\usepackage{tikz}

\usepackage[letterpaper,top=2cm,bottom=2cm,left=3cm,right=3cm,marginparwidth=1.75cm]{geometry}

\newtheorem{theorem}{Theorem}
\newtheorem{corollary}{Corollary}
\newtheorem{example}{Example}
\newtheorem{remark}{Remark}
\def\P{{\mathbf P}}

\providecommand{\keywords}[1]
{
  \small	
  \textbf{\textit{Keywords---}} #1
}
\usepackage{amsmath}

\usepackage[colorlinks=true, allcolors=blue]{hyperref}
\usepackage{color}

\title{ 
Efficient Remote Monitoring through Noisy Random Access with Retransmissions}

\author{Foss S.G.$^{1}$, Kim D.K.$^{2}$ \footnote{This research of Dmitriy Kim and Sergey Foss is funded by the Science
	Committee of the Ministry of Science and Higher Education of the Republic of
	Kazakhstan (Grant No. AP19680230)}, Turlikov A.M.$^{3}$\footnote{This work of Andrey Turlikov  is an output of a research project implemented as part of the Basic Research Program at HSE University}\\
    \small $^{1}$Heriot-Watt University, EH14 4AS, UK (e-mail: s.foss@hw.ac.uk)  \\
    \small $^{2}$Narxoz University, 050035, Kazakhstan (e-mail: dmitriy.kim@narxoz.kz) \\
    \small $^{3}$Higher School of Economics (HSE), HSE Campus in St. Petersburg, 194100, 
Russia (e-mail: atiurlikov@hse.ru)  \\}

\date{}

\begin{document}

\maketitle

\begin{abstract}
We consider a rare event monitoring system consisting of a set of devices and a base station, where devices transmit information about rare events to the base station using a random multiple access scheme.
We introduce a model in which the presence of noise in the multiple access channel can cause message loss even in the absence of transmission collisions. The occurrence of events is modeled by a family of independent two-state Markov chains (with states 0 and 1). We analyze how repeated transmissions affect system performance.
Two efficiency criteria are proposed and studied: the maximum probability that a message about an event from a fixed device is successfully delivered to the base station and the maximum frequency at which the base station successfully receives updates about the entire system. For each criterion, we determine the optimal number of retransmissions as a function of the system parameters.
\end{abstract}

\keywords{
noisy random multiple access channel, slotted ALOHA, Markov sources, remote monitoring, optimal number of retransmissions}

\section{Introduction}
\label{sec1}
The problem of monitoring the state of a set of objects distributed across a geographic area arises in numerous fields of human activity. In the vast majority of cases, such monitoring relies on communication systems consisting of multiple devices transmitting data to a base station via a wireless channel \cite{Lopez,Nguyen, Metia}. Given the typically large number of devices and the rarity of the events being monitored, random multiple access methods have proven to be the most efficient for utilizing the shared communication channel \cite{Riahi}.

Alternative approaches such as Time Division Multiple Access (TDMA) and Frequency Division Multiple Access (FDMA) eliminate transmission conflicts by allocating dedicated channel resources to each device. However, this design results in average delays -- from event occurrence to successful information delivery -- that grow linearly with the number of devices, regardless of the low frequency of events.

Random multiple access communication systems have been studied extensively. A foundational model capturing the key characteristics of such systems was introduced in the seminal works \cite{Tsybakov, Capetanakis}. Following the terminology established in those papers, we refer to the devices as {\it users}.

In the majority of earlier studies, beginning with the seminal works \cite{Tsybakov, Capetanakis}, the focus has been on minimizing transmission delay in systems where a set of users  communicates with a base station (BS). More recently, a shift in problem formulation has been proposed in several papers \cite{RM_1, RM_2, RM_3}, where the objective is to estimate the states of a family of users rather than merely ensuring rapid message delivery.

Let us consider this new formulation -- novel in the context of multiple access systems -- as presented in \cite{RM_1}. That paper builds on the random multiple access model originally introduced by Tsybakov and Capetanakis, specifically in the case of a finite number of users. Each user transmits information about the occurrence of rare events to the BS using a random access protocol. The occurrence of these events is modeled by a collection of independent Markov chains, one per user. The works \cite{RM_1, RM_2, RM_3} investigate various approaches to estimating the states of these chains based on observations of the output from a shared communication channel.

 In \cite{RM_1}, each user operates independently and can be in one of two states -- $0$ or $1$ -- at any given time. The transitions between these states are governed by a homogeneous Markov chain with transition probabilities $p_{i,j},$ $i,j = 0,1$.  Unlike in \cite{Tsybakov, Capetanakis} and many subsequent studies -- where the content of the transmitted information is deemed irrelevant -- \cite{RM_1} demonstrates that accounting for message content can significantly improve system efficiency. In particular, a user may transmit a message over the shared channel that includes both its identifier and its current Markov state.
The model allows for three possible outcomes on the channel: {\bf Empty} (no user transmits), {\bf Conflict} (two or more users transmit simultaneously) and {\bf Success} (exactly one user transmits).

The paper \cite{RM_1} addresses two interrelated components: (1) the user’s transmission policy -- i.e. the decision rule determining when to transmit -- and (2) the BS's estimation policy -- i.e. how to infer the users' states based on the observed sequence of channel outcomes. Several performance criteria for the BS’s state estimation are proposed, and optimal strategies for both the users and the BS are derived with respect to these criteria.

The wide variety of user and BS strategies proposed in \cite{RM_1} and related works makes it challenging to apply their results directly in the design of practical wireless systems for rare event monitoring.
Moreover, the model analyzed in \cite{RM_1} and several similar studies overlooks certain critical aspects of data reception and processing in real-world wireless systems. We propose a modified model, closely related to that of \cite{RM_1}, but incorporating additional practical constraints. In our model, the BS is unable to distinguish between an {\bf Empty} slot and a {\bf Conflict} -- i.e., the BS cannot tell whether the absence of a successful transmission is due to no user transmitting or due to a collision. This reflects the behavior of low-cost wireless technologies, such as LoRa-based systems \cite{Kim, Ali, Garlisi, Ray, KimHwang, Helou}, where such ambiguity is common and significantly complicates the implementation of random multiple access protocols \cite{Foss}.

Furthermore, we assume that even when only a single user transmits, the corresponding message may still fail to be received by the BS with a certain probability -- due to channel noise or other forms of interference. This modification brings the model closer to the behavior of real wireless systems, where message loss can occur independently of contention.

In this paper, we assume that a user can remain in state $1$  for only a very short duration, and we consider the extreme case where $p_{1,0}=1$. Under this assumption, we investigate the potential benefits of retransmissions within the monitoring system. It is worth noting that the idea of employing retransmissions to enhance the performance of random access systems has a long history (see, for example, \cite{Munari, Raza}). Repeated transmissions can reduce delay in systems where the signal propagation time significantly exceeds the message transmission time, as well as in systems with multiple channels \cite{Galinina}. In \cite{Liva}, retransmissions combined with advanced signal processing at the BS are shown to significantly improve the probability of successful message delivery.

In this work, we focus on a simple operational strategy for the users: once a user transits to state $1$, it transmits information about this state to the BS multiple times. We introduce two performance criteria related to the probability of successful message delivery and derive the optimal number of retransmissions as a function of the system parameters.

It is important to note that the successful reception of a message by the BS provides information about the state of the entire system over a certain time interval. Specifically, when the BS receives a message from any user, it learns not only the state of that particular user but also gains indirect information about the other users who are in state $0$. Since users in state $0$ do not transmit, their silence over a sequence of time slots -- corresponding to the duration of the repeated transmissions -- implies that they have remained in state $0$ throughout that interval. Consequently, improving the probability of successful message delivery reduces the overall uncertainty about the system's state by allowing the BS to infer more about the status of all users, not just the transmitting ones.

The paper is organized into six Sections, followed by two Appendices containing the proofs of the main results. In Section 2, we introduce the model with a finite number of users. Section 3 presents the corresponding analytical results. Section 4 extends the framework to a model with an infinite number of users, and the main findings for this case are formulated. In Section 5, we analyze the optimal number of retransmissions based on the proposed efficiency criteria. Section 6 provides several numerical examples that illustrate and support the theoretical results. A summary of the key contributions is presented in the Conclusion Section.

\section{Mathematical model with a finite number of users}

\label{sec2}

We consider a communication network consisting of $N\geq 1$ users designed for monitoring rare events in a distributed system with a single BS. Each user independently transmits messages to the BS over a shared communication channel.

Time is assumed to be slotted, i.e. divided into discrete time intervals (slots), and each message transmission occupies exactly one time slot. If two or more users transmit simultaneously, a {\bf Conflict} occurs, and none of the messages is successfully delivered to the BS.

At time slot $n$, every user can be in one of two states, denoted by $0$ and $1$. State $0$ indicates the absence of a message at time $n$, while state $1$ signifies the appearance of a new message to be sent at time $n$.

We assume that each user independently receives new messages at random time instants. Specifically, at each time slot, a user receives a new message with probability $q\in (0,1)$,  independently of all other users. Once a user receives a message (i.e. transits to state $1$), it cannot receive another message during the same time slot or in the immediately following one.

We denote by $T_n^{(j)}$ the state of the $j$-th user at time $n$. We assume that, for $j=0,1,2,\ldots, N-1$, the sequences $\{T_n^{(j)}, -\infty  < n < \infty\}$ are independent of each other and that, for any fixed $j$, the sequence $T_n^{(j)}$ forms a time-homogeneous 
Markov chain with state space $\{0, 1\}$ and transition probabilities  
\begin{align*}
p_{0,1} & = \P(T_n^{(j)} = 1 \ | \ T_{n-1}^{(j)} = 0) = q,    
\\
p_{0,0} &= \P(T_n^{(j)} = 0 \ | \ T_{n-1}^{(j)} = 0) = 1-q,
\\
p_{1,0} &= \P(T_n^{(j)} = 0 \ | \ T_{n-1}^{(j)} = 1) = 1,  
\\
p_{1,1} &= \P(T_n^{(j)} = 1 \ | \ T_{n-1}^{(j)} = 1) = 0.
\end{align*}
where $q\in (0,1)$ is the probability of the appearance of a new message. This Markov chain is irreducible  and aperiodic, and it possesses a  unique stationary distribution $(\pi_0, \pi_1)$ where 
\begin{align*}
\pi_0=\frac{1}{q+1}, \quad \pi_1=\frac{q}{1+q}.    
\end{align*}
We assume that all users operate in the stationary regime. 

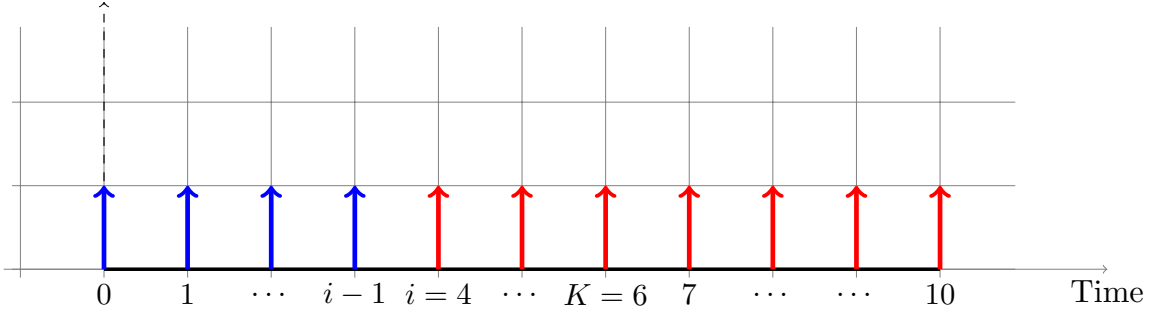
\begin{figure}[ht!]
\centering
\resizebox{\columnwidth}{!}{
    \begin{tikzpicture}[domain=0:4]
    
\draw[very thin,color=gray] (-0.1,-0.1) grid (11.9,2.9);
\draw[->, very thin,color=gray] (-0.2,0) -- (13.0,0) node[ black,below] {Time};
\draw[very thick] (1,0) -- (11,0);
\draw[->, dashed] (1,0) -- (1,3.2);

\draw[->,blue, ultra thick] (1,0) -- (1,1);
\draw[->,blue, ultra thick] (2,0) -- (2,1);
\draw[->,blue, ultra thick] (3,0) -- (3,1);
\draw[->,blue, ultra thick] (4,0) -- (4,1);
\draw[->,red, ultra thick] (5,0) -- (5,1);
\draw[->,red, ultra thick] (6,0) -- (6,1);
\draw[->,red, ultra thick] (7,0) -- (7,1);
\draw[->,red, ultra thick] (8,0) -- (8,1);
\draw[->,red, ultra thick] (9,0) -- (9,1);
\draw[->,red, ultra thick] (10,0) -- (10,1);
\draw[->,red, ultra thick] (11,0) -- (11,1);

\node[scale=0.95] at (1.,-0.3) {$0$};
\node[scale=0.95] at (2.0,-0.3) {$1$};
\node[scale=0.95] at (3.0,-0.3) {$\dots$};
\node[scale=0.95] at (4.0,-0.3) {$i-1$};
\node[scale=0.95] at (5.0,-0.3) {$i=4$};
\node[scale=0.95] at (6.0,-0.3) {$\dots$};
\node[scale=0.95] at (7.0,-0.3) {$K=6$};
\node[scale=0.95] at (8.0,-0.3) {$7$};
\node[scale=0.95] at (9.0,-0.3) {$\dots$};
\node[scale=0.95] at (10.0,-0.3) {$\dots$};
\node[scale=0.95] at (11.0,-0.3) {$10$};

\end{tikzpicture}
}
\caption{An example of overlapping transmissions of two messages from a single user.}
\label{fig_1}
\end{figure}

Assume that, after receiving a new message, a user transmits it to the BS $K+1$ times (where $K\ge 0$ is an integer): the first transmission occurs at the time the message arrives, followed by $K$ consecutive retransmissions in the subsequent time slots.
Further, we assume that if a user receives another new message within $K$ time units of the previous one, it stops transmitting the earlier message and begins transmitting the new one. In the example shown in Figure~\ref{fig_1}, where $K=6$, the user receives two messages within the time interval $[0,10]$: the first at time $0$, and the second at time $i=4$. As a result, the first message is transmitted only $4$ times -- at moments $0,1,2,3$. Starting from time $4$, the user begins transmitting the second message, which is sent a total of $K+1=7$ times, until time $n=10$.

Each transmitted message includes the user’s identification number and the time of message occurrence.
  In Figure \ref{fig_1}, the user with identification number $0\leq j\leq N-1$ transmits  messages with content $(j,0)$ at times $i=0,1,2,3$, and  messages with content $(j,4)$ at times $i=4,5,\dots, 10$.

A transmission at time $n$ is considered {\it successful} if the message is sent in time slot $n$ and no other user transmits in the same slot.

Additionally, we assume that even a successfully transmitted message may be corrupted by random noise at slot $n$, rendering it unreadable by the BS. We say a message is {\it delivered} at time $n$ if it is both successfully transmitted in slot $n$ and not distorted by noise.

Introduce the following events:
\begin{align*}
F_n &= \{\text{a successful transmission occurs in slot $n$}\},
\\
D_n &= \{\text{a delivery of message to BS occurs in slot}  \ \ n\}.
\end{align*}

We assume that, for any $n$, the conditional probability that a message is delivered -- given that it is successfully transmitted -- is equal to
\begin{align}
\label{eps}
\P\left(D_n \ | \ F_n\right) = 1-\varepsilon,
\end{align}
where $\varepsilon\in[0,1)$ is a fixed number.\\
Additionally, we assume that noise events occurring at distinct time slots are conditionally independent, i.e. for any 
 $m$ and for any collection of times $l_0 < l_1 < \ldots < l_{m-1} < l_m$, 
\begin{align}
   \P \left( D_{l_{0}}D_{l_{1}}\dots D_{l_{m}}
    \ \big| \ F_{l_{0}}F_{l_{1}}\dots F_{l_{m}}\right) &=
   \prod_{s=0}^m \P (D_{l_{s}} \ | \ F_{l_{s}}) \nonumber
   \\
   &=(1-\varepsilon)^{m+1}.
 \end{align}

 The objectives of this work are twofold: (i) to compute the probability of successful message delivery to the BS in the presence of noise, and (ii) to investigate how the optimal number of retransmissions $K$ -- which maximizes this delivery probability -- depends on the system  parameters $N$, $q$ and $\varepsilon$. 
 
\section{Results for the finite-user model}
\label{sec3}

In this section, we analyze the probability, denoted by $W$, that exactly  one user receives a new message at time $n=0$ and that this message is both successfully transmitted and delivered to the BS. In steady state, $W$ also represents the long‑run frequency of message deliveries to the BS. We refer to $W$ as the {\it system probability}.

We also consider the conditional probability $V$ of delivery given that a particular user receives a message at $n=0$. We call $V$ the {\it individual probability}.

It follows directly that the system and the individual delivery probabilities differ by the factor $Nq/(1+q)$.
Indeed, in the stationary regime, each user generates a new message in a slot with probability $\frac{q}{1+q}$, and since there are $N$ users, the overall arrival rate of messages is 
$Nq/(1+q)$. 

Now we state our first theorem and its corollaries. The proof is provided in Appendix 1. 

\begin{theorem}
\label{T1}
The individual probability of message delivery is equal to 
\begin{align*}
 V &= (1-\varepsilon)\left(\frac{1}{1+q}\right)^{N-1} (1-q)^{(N-1)K}\times
 \\
&\left[1+\varepsilon (1-q)^{N-1}\frac{1-\varepsilon^K (1-q)^{K N}}{1-\varepsilon (1-q)^{N}}+
\frac{1- \left(1-q\right)^{N-1}}{1-\varepsilon (1-q)^{N}}\times{}
\right.
\\
&\left.
\left( \frac{1-(1-q)^K}{q}- \varepsilon(1-q)^{N+K-1} \frac{1-\varepsilon^K(1-q)^{(N-1)K}}{1-\varepsilon(1-q)^{N-1}}\right)\right], 
\end{align*}
and the system probability $W$ is
\[
W = N \frac{q}{1+q} V.
\]

\end{theorem}

\begin{corollary}
    If $\varepsilon =0$, then 
    \begin{align}
    \label{V_finit}
    V &= \left(\frac{1}{1+q}\right)^{N-1} (1-q)^{(N-1)K}\times{} \nonumber
    \\
    &\left(1+\left(1- \left(1-q\right)^{N-1}\right)  \frac{1-(1-q)^K}{q}\right)     
    \end{align}
     and 
    \begin{align*}
    W &= N \frac{q}{1+q}\left(\frac{1}{1+q}\right)^{N-1} (1-q)^{(N-1)K} \times{}
    \\
    &\left(1+\left(1- \left(1-q\right)^{N-1}\right)  \frac{1-(1-q)^K}{q}\right).    
    \end{align*}
\end{corollary}

It follows from the above formulae that for $\varepsilon =0$ the individual and the system probabilities are non-increasing functions of the variable $K\geq 0$ that may be represented as  
\begin{align*}
c_1 (1-q)^{(N-1)K}\left(c_2+c_3(1-q)^{K}\right),    
\end{align*}
where $c_1, c_2, c_3$ are some constants. Therefore, we arrive at the following conclusion. 

\begin{corollary}
 If $\varepsilon =0$ and $N>1$, then the values of $V$ and $W$ are maximal for $$K=0.$$
\end{corollary}

\begin{remark}
Because the factor $Nq/(1+q)$ does not depend on $K$, it follows that, for any $\varepsilon$, the values of $K$ that maximize the system delivery probability and the individual delivery probability coincide.
\end{remark}

\begin{remark}
If a message is delivered to the BS, then necessarily all other users must have been in the inactive state $0$ not only at the delivery time but also at each of the preceding $K$ slots. Consequently, a successful delivery provides perfect information about the system’s state over a window of $K+1$  slots, and choosing the value of $K$ that maximizes the delivery probability minimizes the system’s uncertainty.
\end{remark}

\begin{remark}
\label{rem_3}
Now assume that each transmitted message also includes information on the user’s states during the preceding $K$ time slots. Under this assumption, transmissions are never preempted: every transition to state $1$ generates exactly $K+1$ transmissions. To distinguish these performance measures from those in Theorem \ref{T1}, we denote the individual and system delivery probabilities here by $\widetilde{V}$ and
$\widetilde{W}$. The individual delivery probability then takes the form
\begin{align}
\label{V_tilda}
\widetilde{V} &= (1-\varepsilon)\frac{(1-q)^{(N-1)K}}{\left(1+q\right)^{N-1}}\times{}
\nonumber
\\
&\left(
\frac{1-\varepsilon^{K+1} (1-q)^{(N-1)(K+1)}}{1-\varepsilon (1-q)^{(N-1)}}+\frac{1-(1-q)^{N-1}}{1- \varepsilon(1-q)^{N-1}}\times{}
\right. \nonumber
\\
&\left.
\left(K- \varepsilon (1-q)^{N-1} \frac{1-\left(\varepsilon(1-q)^{N-1}\right)^{K}}{1- \varepsilon(1-q)^{N-1}}\right)\right)
\end{align}
and the system probability $\widetilde{W}$ is 
\[
\widetilde{W} = N \frac{q}{1+q} \widetilde{V}.
\]
These conclusions are easily obtained from the proof of Theorem $\ref{T1}$ by removing the multiplier $(1-q)^{i+j-1}$ in the formula $(\ref{thu})$. 

The user’s transmission of its state history over the previous $K$ time slots can be represented by an additional binary vector of length $K+1$. Previously, each message carried the user’s identifier $j\in \{0,1,\ldots,N-1\}$ and the time of appearance (e.g., $n=0$). Under the new scheme, the message additionally includes a binary vector of length $K+1$, where each entry records its state in the corresponding past slot.
\end{remark}

\begin{figure}[ht!]
\centering
\resizebox{\columnwidth}{!}{
    \begin{tikzpicture}[domain=0:4]
\draw[very thin,color=gray] (-0.1,-0.1) grid (11.9,2.9);
\draw[->, very thin,color=gray] (-0.2,0) -- (13.0,0) node[ black,below] {Time};
\draw[very thick] (1,0) -- (11,0);
\draw[->, dashed] (1,0) -- (1,3.2);

\draw[->,blue, ultra thick] (1,0) -- (1,1);
\draw[->,blue, ultra thick] (2,0) -- (2,1);
\draw[->,blue, ultra thick] (3,0) -- (3,1);
\draw[->,blue, ultra thick] (4,0) -- (4,1);
\draw[->,blue, ultra thick] (4.9,0) -- (4.9,1);
\draw[->,red, ultra thick] (5.1,0) -- (5.1,1);
\draw[->,blue, ultra thick] (5.9,0) -- (5.9,1);
\draw[->,red, ultra thick] (6.1,0) -- (6.1,1);
\draw[->,blue, ultra thick] (6.9,0) -- (6.9,1);
\draw[->,red, ultra thick] (7.1,0) -- (7.1,1);
\draw[->,red, ultra thick] (8,0) -- (8,1);
\draw[->,red, ultra thick] (9,0) -- (9,1);
\draw[->,red, ultra thick] (10,0) -- (10,1);
\draw[->,red, ultra thick] (11,0) -- (11,1);

 \node[scale=0.95] at (1.,-0.3) {$0$};
 \node[scale=0.95] at (2.0,-0.3) {$1$};
 \node[scale=0.95] at (3.0,-0.3) {$\dots$};
 \node[scale=0.95] at (4.0,-0.3) {$i-1$};
 \node[scale=0.95] at (5.0,-0.3) {$i=4$};
 \node[scale=0.95] at (6.0,-0.3) {$\dots$};
 \node[scale=0.95] at (7.0,-0.3) {$K=6$};
 \node[scale=0.95] at (8.0,-0.3) {$7$};
 \node[scale=0.95] at (9.0,-0.3) {$\dots$};
 \node[scale=0.95] at (10.0,-0.3) {$\dots$};
 \node[scale=0.95] at (11.0,-0.3) {$10$};
\end{tikzpicture}
}
\caption{An illustrative example of two overlapping transmissions from the same user, where each message conveys a binary vector of length $K+1$ encoding the user’s state over the previous $K$ time slots.}
\label{fig_3}
\end{figure}
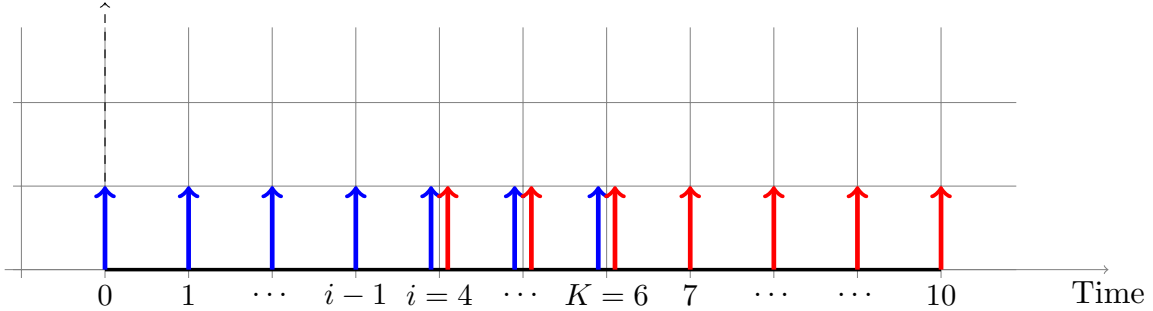

To illustrate Remark $\ref{rem_3}$, we provide another example, see Fig. $\ref{fig_3}$. 
 
Suppose $K=6$. A message transmitted at time $n=0$ might contain the vector $(1000000)$
indicating that the user is active at $n=0$ and inactive during the previous six slots. At $n=1$, the user would send 
$(0100000)$, at $n=2$ would send $(0010000)$, and at $n=3$ would send $(0001000)$. 
If at $n=4$ the user both receives a new message and was active four slots earlier, the transmitted vector would be
$(1000100)$, 
signifying “active at $n=0$” (first entry) and “active at $n=4$” (fifth entry).

\section{Model with an infinite number of users}
\label{sec5}

The expressions for the probabilities of successful transmission in Theorem \ref{T1} simplify considerably in the infinite‑user limit. Specifically, letting $N\to\infty$ and $q\to 0$ such that $Nq\to\lambda \in (0,\infty)$  yields the classical Poisson approximation.
In this limit, the system reduces to a traditional ALOHA random multiple-access model augmented with retransmissions (i.e., repeated message transmissions, as per our terminology).
Because $q\to 0$, the chance that any individual user generates two messages within a finite interval vanishes, yielding a one‑to‑one mapping between messages and users. 

Let $\xi_n$ be the number of messages arriving
in the system at time $n$. It follows from our assumptions that $\{\xi_n\}_{n=-\infty}^\infty$ form a sequence of independent random variables having the same Poisson distribution with parameter $\lambda$.

We continue to assume that any new message undergoes $K+1$ transmission attempts: a message arriving at time $n$ is transmitted in time slots $n,n+1,\ldots,n+K$.

We retain the terminology “successfully transmitted” and “delivered” as defined previously. We continue to assume that noise events in different time slots are conditionally independent, and we keep the same notation $F_n$, $D_n$, the accompanying assumptions (\ref{eps}).
Applying the limiting transition in the statement of Theorem \ref{T1},  
\begin{align*}
N \to \infty,\ q\to 0,    
\end{align*}
with
\begin{align}
\label{Nq}
Nq \to \lambda  \ (\text{and therefore} \  N\frac{q}{1+q}\to\lambda),    
\end{align}
 we may obtain an analogous result for the model with an infinite number of users. 

To avoid confusion with the finite‑user notation from Theorem \ref{T1}, we denote the corresponding probabilities in the infinite‑user model by $V_{\infty}$ and $W_{\infty}$.

\begin{theorem}
\label{T2}
In the model with an infinite number of users, the individual probability of the message delivery is equal to
\begin{align}
\label{V_inf}
V_\infty &= \frac{1-\varepsilon}{1-\varepsilon e^{-\lambda}} e^{-(K+1)\lambda} \left((1-e^{-\lambda}) (K+1) +\right.\nonumber
\\
&\left.
\frac{1-\varepsilon}{1-\varepsilon e^{-\lambda}} e^{-\lambda} 
\left(1- \left(\varepsilon e^{-\lambda}\right)^{K+1}\right) \right) 
\end{align}
and the system probability is  
\begin{align*}
W_\infty = \lambda V_\infty.    
\end{align*}
\end{theorem}

\begin{remark}
Analogously to the case of a finite number of users, we may conclude that the individual and the system delivery probabilities differ by the factor $\lambda$ that appears at the limiting transition (see  (\ref{Nq})): 
\begin{align*}
N \frac{q}{1+q}\to\lambda    
\end{align*}
 and does not depend on $K$.
   Consequently, both delivery probabilities attain their maximum at the same value of $K$.
\end{remark}

\begin{remark} 
We derived Theorem  \ref{T2} by applying the formal limiting transition from Theorem \ref{T1}. In Appendix 2, we also outline a direct proof of Theorem \ref{T2}.
\end{remark}

\begin{remark} 
We should also note that the marginal probabilities of message delivery remain unchanged under the assumption that a user interrupts the transmission of a previously received message upon the arrival of a new one (see Remark~\ref{rem_3}). Indeed, in the limiting regime considered here, the probability that two messages are received by the same user within any finite time interval is zero.
\end{remark}

\begin{corollary}
 If $\varepsilon =0$, then
 \begin{align*}
 V_\infty = e^{-(K+1)\lambda} \left(1+ K\left(1-e^{-\lambda}\right)\right),    
 \end{align*}
 \begin{align*}
 W_\infty = \lambda e^{-(K+1)\lambda} \left(1+ K\left(1-e^{-\lambda}\right)\right).    
 \end{align*}
\end{corollary}

Clearly, for $\varepsilon =0$,  the probabilities $V_\infty$ and $W_\infty$ are non-increasing functions of parameter $K\geq 0$, so the following holds: 

\begin{corollary}
 If $\varepsilon = 0$ and $\lambda > 0$, then both the individual and the system delivery probabilities, $V_\infty$ and $W_\infty$, attain their maximum values at $K = 0$.
\end{corollary}

\section{Determining optimal values of $K$}
\label{sec6}

The results of Theorem~\ref{T2} for the case of an infinite number of users allow us to analyze the optimal number of retransmissions $K$ that maximizes the probability of message delivery. Studying this limiting case simplifies the analysis and provides useful insights into the behavior of the finite-user model. Moreover, transferring conclusions from the infinite-user setting to the finite case yields approximations that are often close to optimal. Therefore, we focus here on the model with an infinite number of users.

In this setting, for any fixed value of $K$, the model is fully characterized by two parameters: the message arrival rate $\lambda$ and the noise level $\varepsilon\in (0,1)$. Using expression~\eqref{V_inf}, we evaluate the delivery probability for each integer $K\ge 0$. For each such value of $K$, we define the {\it optimal region} in the $(\varepsilon, \lambda)$ plane as the set of points for which $V_{\infty}$, as given by~\eqref{V_inf}, is maximized at that particular $K$.

\begin{figure}
 \centering
  \includegraphics[width=12cm]{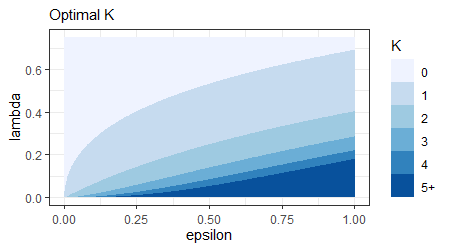} 
  \caption{ Partitioning the domain 
   $\varepsilon\in (0,1)$, $\lambda \in (0,0.75)$ into optimal  regions for the values of $K\in \{0,1,2,3,4\}$ and for the union of regions for $K\geq 5$.}
  \label{fig_4}
  \end{figure}

In Figure~\ref{fig_4}, we illustrate these {\it optimal regions} for $K=0,1,2,3,4$, as well as the combined region corresponding to $K\geq 5$. It can be shown that for values of $\varepsilon$ sufficiently close to $1$, the point with coordinates $(\varepsilon, \lambda =1/(K+1))$ lies within the optimal region for $K$ retransmissions.

  Using the results derived for the limiting case, we can determine the optimal value of $K$ for any given set of model parameters $N>1, 0<q<1$ and $\varepsilon <1$. To do so, it is sufficient to identify the optimal region -- defined in terms of retransmissions $K$ -- that contains the point with coordinates $(\varepsilon,\lambda = Nq/(1+q))$. As shown in Figure~\ref{fig_4}, the optimal value of $K$ increases as $\lambda$ decreases, for fixed $\varepsilon >0$. Likewise, for fixed $\lambda$, the optimal number of retransmissions increases with increasing $\varepsilon$. Thus, repeated transmissions are most beneficial when $Nq/(1+q)$ is small and the distortion probability $\varepsilon$ is high.
    If $\varepsilon$ is close to 1, the solution of the linear equation $(x+1)Nq/(1+q)=1$, rounded to the nearest integer, can be used as an approximation for the optimal value of $K$.

Let us now derive more accurate quantitative conclusions based on expression $(\ref{V_inf})$.
This expression enables a straightforward analysis, allowing us to approximate the optimal value of $K$ that maximizes the probability of message delivery $V$. 

Consider an auxiliary function
\begin{align*}
g(x) =e^{-x\lambda} \left( (1-e^{-\lambda})x+
e^{-\lambda}\frac{1-\varepsilon}{1-\varepsilon e^{-\lambda}} \left(1-(\varepsilon e^{-\lambda})^{x}\right)\right)    
\end{align*}
and find the value of $x$ that maximizes its value. The obtained value is close to the optimal $K+1$ in the formula $(\ref{V_inf})$. Let us find the derivative of the function $g$ and equate it to zero
\begin{align*}
-\lambda \left((1-e^{-\lambda})x+
e^{-\lambda}\frac{1-\varepsilon}{1-\varepsilon e^{-\lambda}} \left(1-(\varepsilon e^{-\lambda})^{x}\right)\right) 
\\    
+ 1-e^{-\lambda} +e^{-\lambda}\frac{1-\varepsilon}{1-\varepsilon e^{-\lambda}}(\lambda - \log\varepsilon)(\varepsilon e^{-\lambda})^{x}=0.
\end{align*}

After a minor simplification we obtain that the required value is the root to the equation
\begin{align}
\label{Fx0}
F(x) =0,
\end{align}
where
\begin{align}
\label{F}
F(x) &= x- 1/\lambda + A\left(1 -\left(1 - \lambda^{-1}\log(\varepsilon e^{-\lambda})\right)(\varepsilon e^{-\lambda})^{x}\right)
\end{align}
and 
\begin{align*}
A = \frac{1-\varepsilon}{1-\varepsilon e^{-\lambda}} \cdot \frac{e^{-\lambda}}{1-e^{-\lambda}} >0.    
\end{align*}
Note that $F(0) <0$ and that $F(x)$ increases and tends to infinity as $x$ grows. Therefore,
equation (\ref{Fx0}) has a unique solution $x^*$ on the positive halfline.

It is not difficult to check that the functions $F'(x)$ and $F''(x)$ are sign-constant.  
Therefore, we can use the Newton-Raphson method, which provides a high convergence rate for any non-negative initial value $x_0$:
\begin{align*}
x_{n+1} = x_n - \frac{F(x_n)}{F'(x_n)},    
\end{align*}
where
\begin{align}
\label{dF}
F'(x) = 1+ A\left(2 - \frac{\log\varepsilon}{\lambda}\right) (\lambda - \log\varepsilon) (\varepsilon e^{-\lambda})^{x}.    
\end{align}

 To obtain an explicit approximation, we can choose an initial value $x_0$ and select the first iteration $x_1$:
 \begin{align*}
 x^* = x_0 - \frac{F(x_0)}{F'(x_0)} +\alpha(\lambda, \varepsilon),    
 \end{align*}
 where $F(x)$ and $F'(x)$ are defined in the formulas $(\ref{F})$ and $(\ref{dF})$. The error $\alpha(\lambda, \varepsilon)$ is of order $O((x_0-x^*)^2)$. More precise estimates of the error require cumbersome computations and are therefore omitted. Thus, to increase the accuracy, it is desirable to choose the initial value of $x_0$ close to the sought $x^*$. 
 For example, if $\lambda$ is small and $\varepsilon$ is close to 1, the optimal $K$ behaves like $1/{\lambda} -1$ (see $(\ref{V_inf})$) and it is better to choose
\begin{align*}
x_0 =1/\lambda.    
\end{align*}
Then 
\begin{align*}
x^* \approx\frac{1}{\lambda} - \frac{F\left(1/\lambda\right)}{F'\left(1/\lambda\right)}.
\end{align*}

Once the representation for $x^*$ has been chosen, one can use the approximation 
\[
K +1\approx x^*.
\]

\section{Examples}
\label{sec7}

First, we consider the limiting case where the system has an unlimited number of users, and we set $\varepsilon = 0.4$ and $\lambda = 0.02$. Note that the point with these coordinates lies in the region corresponding to $K \geq 5$ (see Fig.\ref{fig_4}). Using formula$(\ref{V_inf})$, we compute the individual delivery probabilities $V_\infty$ for various values of $K$.

For clarity, instead of analyzing the delivery probability $V_\infty$, we examine the probability of message non-delivery, which is given by $1 - V_\infty$. We then plot the dependence of $1 - V_\infty$ on $K$.

Next, we consider a contrasting case where the number of users is finite: let $N = 2$ and choose $q$ such that $\lambda = N \frac{q}{1+q}$. Using the same approach and formula~$(\ref{fig_4})$, we plot, in the same graph, the dependence of the non-delivery probability $1 - V$ on $K$ for this finite-user model.

\begin{example}
\label{ex_1}
Let $\varepsilon = 0.4$, $N = 2$, $\lambda = 0.02$
and 
$
q = \frac{\lambda}{N - \lambda} \approx 0.01.
$

  \begin{figure}[ht!]
      \centering
\includegraphics[width = 12cm]{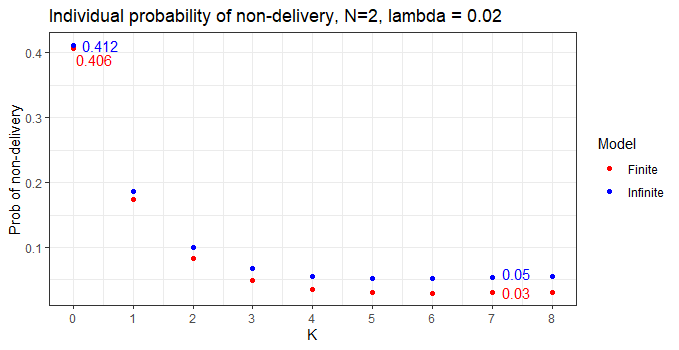}
      \caption{Dependence of the message non-delivery probability on the number of retransmissions $K$ for two scenarios: an infinite number of users (blue curve) and a finite number of users with $N=2$ (red curve), where $\lambda = 0.02$ and $N q/(1+q) = \lambda$.}
      \label{eps0.4l0.02}
  \end{figure}    

From Fig.~\ref{eps0.4l0.02}, it follows that for the selected values of $\lambda$ and $\varepsilon$, the probability of message non-delivery can be reduced by an order of magnitude by choosing the optimal number of retransmissions, $K=7$, compared to the case without retransmissions.
\[
\frac{1-V_{\infty}(7)}{1-V_{\infty}(0)} = 7.9, \quad \frac{1-V(7)}{1-V(0)} = 13.6, 
\]
where$V(0),$ $V_{\infty}(0),$ $V(7),$ $V_{\infty}(7)$ are the individual probabilities of delivery from formulae $(\ref{V_finit})$ and $(\ref{V_inf})$ for $K=0$ and $K=7$, 
\[
1-V_{\infty}(0) = 0.4119, \ 1-V_{\infty}(7) = 0.0521, 
\]
\[
1-V(0) = 0.406, \ 1-V(7) = 0.0298. 
\]
\end{example}

It is noteworthy that the optimal number of retransmissions is the same in both the limiting case of an infinite number of users and the finite case with only $N=2$ users. This observation suggests the conjecture that if the optimal value of $K$ coincides for the infinite-user case and the two-user case, then the same value of $K$ is likely to remain optimal for any finite number of users.

\begin{example}
\label{ex_2}
If we set $N = 10$ under the conditions of the previous example, the graphs of the non-delivery probabilities for different values of $K$ in the finite and infinite user models become practically indistinguishable (see Fig.~\ref{eps0.4l0.02N10}).
\begin{figure}[ht!]
      \centering
\includegraphics[width = 12cm]{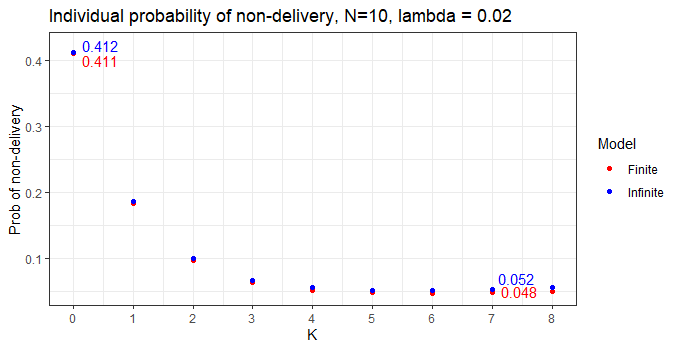}
\caption{Dependence of the non-delivery probability on the number of retransmissions $K$ for the cases of an infinite number of users (blue) and a finite number of users (red), with parameters $N = 10$, $\lambda = 0.02$, and $Nq/(1+q) = \lambda$.}
      \label{eps0.4l0.02N10}
  \end{figure} 
\end{example}

Then, as in Example~\ref{ex_1}, we construct similar graphs for the case $N = 2$, $\lambda = 0.005$, and $\varepsilon = 0.3$.

\begin{example}
\label{ex_3}
Let $\varepsilon = 0.3$, $N = 2$, $\lambda = 0.005$
and 
$
q = \frac{\lambda}{N - \lambda} \approx 0.0025.
$

  \begin{figure}[ht!]
      \centering
\includegraphics[width = 12cm]{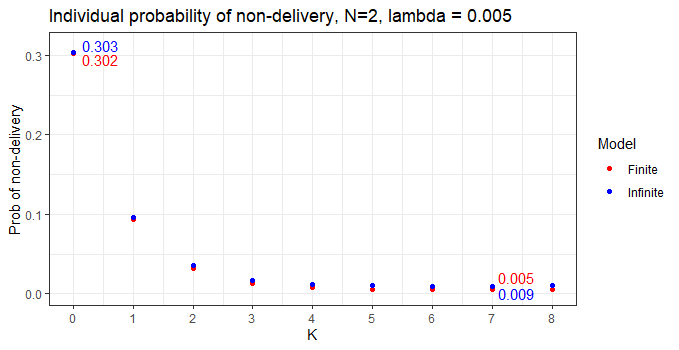}
      \caption{Dependence of the probability of non-delivery on the number of retransmissions $K$ for the cases of an infinite number of users (blue) and a finite number of users (red), with $N = 2$, $\lambda = 0.005$, and $Nq/(1+q) = \lambda$.}
      \label{eps0.3l0.005}
  \end{figure}

  From the dependencies shown in Fig.~\ref{eps0.3l0.005}, it follows that for the given values $N = 2$, $\lambda = 0.005$, and $Nq/(1+q) = \lambda$, the use of retransmissions can reduce the probability of non-delivery by up to two orders of magnitude:
\[
\frac{1-V_{\infty}(8)}{1-V_{\infty}(0)} = 30.76, \quad \frac{1-V(8)}{1-V(0)} = 58, 
\]
\[
1-V_{\infty}(0) = 0.3035, \ 1-V_{\infty}(7) = 0.0098, 
\]
\[
1-V(0) = 0.3017, \ 1-V(7) = 0.0053. 
\]
\end{example}

Let us now compare the values of $1 - V$ and $1 - \widetilde{V}$ (see formulas~(\ref{V_finit}) and~(\ref{V_tilda})), which represent the individual message non-delivery probabilities in two scenarios: when the message contains information only about the current transition to the active state, and when it additionally includes the user’s state during the previous $K$ time instants (see Remark~\ref{rem_3}).

\begin{example}
\label{ex_4}
Let $\varepsilon = 0.4$, $N = 2$, and $q = 0.01$. A graphical comparison of the message non-delivery probabilities, $1 - V$ and $1 - \widetilde{V}$ (see formulas~(\ref{V_finit}) and~(\ref{V_tilda})), is presented in Figure~\ref{eps0.4q0.01}. The case where the message includes only information about the current activation (corresponding to $V$) is labeled as “$K\text{-}$”, while the case where the message additionally contains information about the user’s state over the previous $K$ time instants (corresponding to $\widetilde{V}$) is labeled as “$K$”.
\begin{figure}[ht!]
      \centering
\includegraphics[width = 12cm]{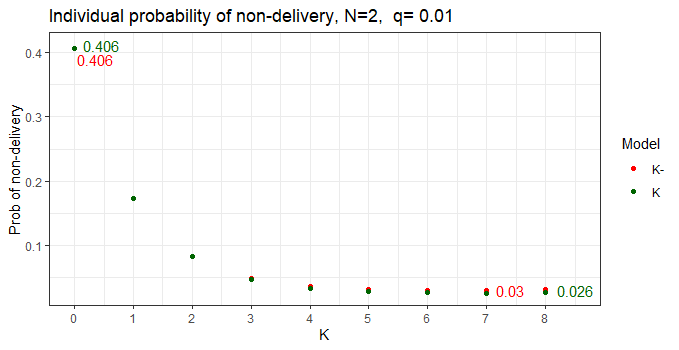}
      \caption{Dependence of the non-delivery probabilities $1-V$ and $1-\widetilde{V}$ 
 on the number of retransmissions $K$
      for $\varepsilon = 0.4$, $N=2$, $q=0.01$. 
      }
      \label{eps0.4q0.01}
  \end{figure}  

  The figure shows that the non-delivery probability $1 - \widetilde{V}$, corresponding to the case where the message includes information about the user's state over the previous $K$ time instants, is higher than $1 - V$, which corresponds to messages reporting only the current transition to the active state. Moreover, the minimum values of these probabilities are attained at different values of $K$. However, under the given parameters, the difference between the two scenarios is relatively small. Therefore, we now consider alternative parameter values that more clearly highlight the difference between $1 - \widetilde{V}$ and $1 - V$ across various values of $K$.
\end{example}

\begin{example}
Let $\varepsilon = 0.99$, $N = 2$, and $q = 0.00526$. A graphical comparison of the non-delivery probabilities $1 - V$ and $1 - \widetilde{V}$ is presented in Figure~\ref{eps0.99q0.0526}. In the figure, the curve corresponding to the standard message format (transmitting only the current active state) is labeled as “$K\text{-}$” (red), while the curve representing the extended message format (including the user's state over the previous $K$ time instants) is labeled as “$K$” (green).

\begin{figure}[ht!]
      \centering
\includegraphics[width = 12cm]{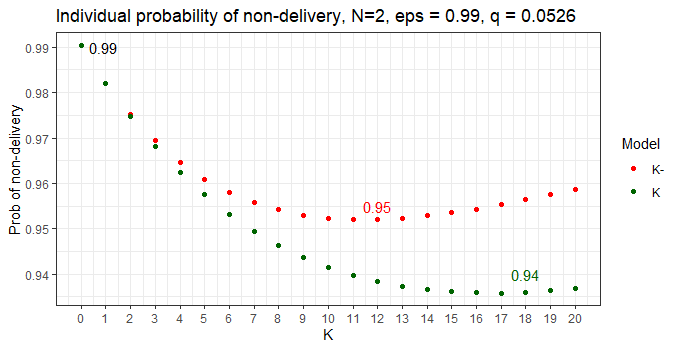}
      \caption{Dependence of the non-delivery probabilities $1 - V$ and $1 - \widetilde{V}$ on the number of retransmissions $K$ for the parameters $\varepsilon = 0.99$, $N = 2$, and $q = 0.0526$.}
      \label{eps0.99q0.0526}
  \end{figure}

The graph above demonstrates that, in the case where additional information about the user's state is transmitted, the probability of non-delivery $1 - \widetilde{V}$ can be significantly lower than $1 - V$, with the optimal number of retransmissions $K$ differing between the two scenarios. However, this effect becomes evident only in extreme (degenerate) cases, such as when $\varepsilon = 0.99$, where the system is nearly non-operational. In more realistic settings -- such as those examined in Examples \ref{ex_1}–\ref{ex_4} -- including information about the user’s state over the previous $K$ time slots does not substantially reduce the probability of non-delivery.

\end{example}

\section{Conclusion}
\label{sec8}

In this paper, we consider a model with $N$ users transmitting messages to a base station (BS) over a slotted multiple access channel. Each user can be in one of two states: active or passive. A passive user transitions to the active state in the next time slot with probability $q$, and remains passive with probability $1-q$. The transition from the active to the passive state occurs deterministically in the following time slot. Upon becoming active, the user transmits a message over the channel $K+1$ times. Each message includes the user's identifier and the time of transition to the active state. 

If two or more users transmit simultaneously in the same time slot, the BS is unable to receive any of their messages due to a collision. We consider a noisy random multiple access channel, where, if only one user transmits in a slot, the BS successfully receives the message with probability $1-\varepsilon$, and fails to receive it with probability $\varepsilon$. The special case $\varepsilon =0$ corresponds to a \emph{noiseless  channel},
in which message loss occurs only due to transmission conflicts. 

Two efficiency criteria are considered: 
probability of delivery to the BS of information about the transition to the active state of some selected user and the frequency of updating information about the system state. The latter is understood as the frequency of delivery of new messages.

An approach is proposed for determining the optimal number of retransmissions $K$ based on the specified system parameters $N$ and $q$, with respect to the considered efficiency criteria. It is shown that the optimal values of $K$ corresponding to both criteria coincide.

Numerical examples are provided to illustrate that, for different parameter sets $(N, q)$ yielding the same value of $Nq/(1+q)$, the corresponding optimal values of $K$ are close.

In the case of a noiseless channel ($\varepsilon = 0$), unsuccessful delivery (non-delivery)  can occur only due to transmission conflicts. Under these conditions, the optimal value of $K$ for the considered performance criteria is zero, implying that retransmissions do not improve efficiency. However, as shown in this paper, the reception of a message from a user provides the BS with information not only about the current time slot but also about the $K$ preceding slots. Taking this into account, it can be conjectured that for alternative performance 
criteria -- such as the fraction of time slots during which the BS possesses complete information about the system -- the use of retransmissions may yield benefits even in a noiseless channel. Exploring such alternative criteria remains an open direction for future research.

The results of this work can be applied in the design and optimization of wireless systems intended for monitoring rare events.

\section{Appendix 1: Proof of Theorem  \ref{T1}}

 Let the users be indexed from $0$ to $N-1$. We focus on determining the individual probability of successful delivery for a specific newly generated message, assuming that it was generated by user $0$ at time $n=0$. For definiteness, we refer to user $0$ as the {\it selected} or {\it individual} user, while all other users (with indices $1$ through $N-1$) will be referred to as {\it other} users.

For a message to be delivered to the BS, it must be successfully transmitted at least once. Recall that the selected user begins transmitting the message to the BS immediately upon activation and continues to do so a total of $K+1$ times. During this retransmission interval, the message may be successfully received by the BS multiple times, potentially resulting in multiple deliveries.

\begin{figure}[ht!]
    \centering
\resizebox{\columnwidth}{!}{
\begin{tikzpicture}[domain=0:1]
\draw[very thin,color=gray] (-0.1,-0.1) grid (15.9,7.9);
\draw[->, very thin,color=gray] (-0.2,0) -- (15.2,0) node[ black,below] {Time};
\draw[->] (5,0) -- (14,0);
\draw[->, dashed] (5,0) -- (5,8.2); 

\draw [black, ultra thick] (0,2) -- (9,2);

\draw[->,blue, ultra thick] (5,0) -- (5,1);
\draw[->,blue, ultra thick] (6,0) -- (6,1);
\draw[->,blue, ultra thick] (7,0) -- (7,1);
\draw[->,blue, ultra thick] (8,0) -- (8,1);
\draw[->,blue, ultra thick] (9,0) -- (9,1);
\draw[->,blue, ultra thick] (10,0) -- (10,1);
\draw[->,blue, ultra thick] (11,0) -- (11,1);
\draw[->,blue, ultra thick] (12,0) -- (12,1);
\draw[->,blue, ultra thick] (13,0) -- (13,1);
\draw[->,blue, ultra thick] (14,0) -- (14,1);

\draw (5.5,1.5) node{$C$};
\draw (6.5,1.5) node{$C$};
\draw (7.5,1.5) node{$C$};
\draw (8.5,1.5) node{$C$};
\draw (9.5,1.5) node{$C$};
\draw (10.5,1.5) node{$S$};
\draw (11.5,1.5) node{$S$};
\draw (12.5,1.5) node{$S$};
\draw (12.5,1.5) node{$S$};
\draw (10.5,2.5) node{$\overline{D}_i$};
\draw (11.5,2.5) node{$\overline{D}_{\dots}$};
\draw (12.5,2.5) node{$D_{i+j}$};
\draw (13.5,2.5) node{$\overline{D}_{\dots}$};
\draw (14.5,2.5) node{$D_K$};

\draw[->, ultra thick, color=cyan] (12.5,3) -- (12.5,7.5);
\draw[->, ultra thick, color=cyan] (14.5,3) -- (14.5,7.5);

\node[scale=0.95] at (5.,-0.3) {$0$};
\node[scale=0.95] at (9.0,-0.3) {$i-1$};
\node[scale=0.95] at (10.0,-0.3) {$i$};
\node[scale=0.95] at (11.0,-0.3) {$\dots$};
\node[scale=0.95] at (12.0,-0.3) {$i+j$};
\node[scale=0.95] at (13.0,-0.3) {$\dots$};
\node[scale=0.95] at (14.0,-0.3) {$K$};

\node[scale=0.95] at (-0.5, 0) {$0$};

\node[scale=0.95] at (-0.5, 2) {$1$};
\node[scale=0.95] at (-0.5, 4) {$2$};
\draw [black, ultra thick] (0,4) -- (3,4);
\node[scale=0.95] at (-0.5, 5) {$\dots$};
\node[scale=0.95] at (-0.5, 6) {$N-1$};
\draw [black, ultra thick] (0,6) -- (8,6);

\draw[->,blue, ultra thick] (0,2) -- (0,3);
\draw[->,blue, ultra thick] (1,2) -- (1,3);
\draw[->,blue, ultra thick] (2,2) -- (2,3);
\draw[->,blue, ultra thick] (3,2) -- (3,3);
\draw[->,blue, ultra thick] (4,2) -- (4,3);
\draw[->,blue, ultra thick] (5,2) -- (5,3);
\draw[->,blue, ultra thick] (6,2) -- (6,3);
\draw[->,blue, ultra thick] (7,2) -- (7,3);
\draw[->,blue, ultra thick] (8,2) -- (8,3);
\draw[->,blue, ultra thick] (9,2) -- (9,3);

\draw[->,blue, ultra thick] (0,4) -- (0,5);
\draw[->,blue, ultra thick] (1,4) -- (1,5);
\draw[->,blue, ultra thick] (2,4) -- (2,5);
\draw[->,blue, ultra thick] (3,4) -- (3,5);

\draw[->,blue, ultra thick] (0,6) -- (0,7);
\draw[->,blue, ultra thick] (1,6) -- (1,7);
\draw[->,blue, ultra thick] (2,6) -- (2,7);
\draw[->,blue, ultra thick] (3,6) -- (3,7);
\draw[->,blue, ultra thick] (4,6) -- (4,7);
\draw[->,blue, ultra thick] (5,6) -- (5,7);
\draw[->,blue, ultra thick] (6,6) -- (6,7);
\draw[->,blue, ultra thick] (7,6) -- (7,7);
\draw[->,blue, ultra thick] (8,6) -- (8,7);

\end{tikzpicture}
}
\caption{The process of transmitting a message (generated at time $n = 0$) by an individual user. During the initial time slots $n = 0, 1, \ldots, i-1$, the transmission is disrupted due to interference from other messages. Beginning at time $n = i$, the message is successfully transmitted in slots $n = i, i+1, \ldots, K$. The message is delivered to the BS twice: the first delivery occurs at time $n = i + j$, and the second at time $n = K$.}
\label{fig_2}
\end{figure}
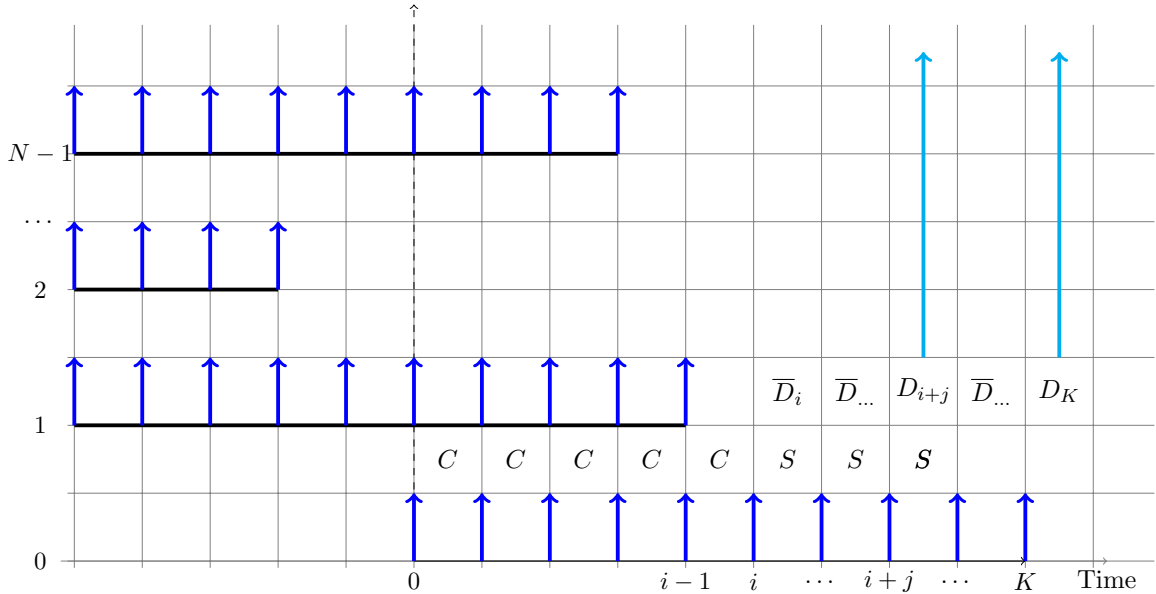

An example illustrating this scenario -- featuring multiple successful transmissions and two distinct deliveries -- is shown in Fig.~\ref{fig_2}.

Recall that the selected message will be delivered to the BS within $K+1$ time slots if it is successfully transmitted at least once and not distorted by random noise. Therefore, the probability of delivery can be expressed as
\[
\P\left(\bigcup_{i=0}^K F_iD_i\right).
\]
Using the classical formula for the probability of the union of events, we get 
\begin{align*}
\P\left(\bigcup_{i=0}^K F_iD_i\right) &=
\sum_{i=0}^K \P(F_iD_i) - \sum_{0\leq i<j\leq K} \P(F_iD_iF_jD_j) + \dots
\\
&
+(-1)^K \P(F_0D_0F_1D_1\dots F_KD_K). 
\end{align*}
Due to the conditional independence (see (\ref{eps})) of the events $D_i$, $i=1,2,\dots, K$, the following equations hold: 

\begin{align}
 \label{incl}
 \P\left(\bigcup_{i=0}^K F_iD_i\right) &= (1-\varepsilon)\sum_{i=0}^K \P(F_i)  
 \nonumber
 \\
 -&(1-\varepsilon)^2\sum_{0\leq i\leq K-1, i< j\leq K} \P(F_iF_{i+j}) 
 \nonumber
 \\
 +&(1-\varepsilon)^3\sum_{0\leq i\leq K-2, i< j< l\leq K} \P(F_iF_lF_j)- \dots 
 \nonumber
 \\
 &+ (-1)^K (1-\varepsilon)^{K+1}\P(F_0F_1\dots F_K).
 \end{align}

The event $F_i F_{i+j}$, with $i + j \leq K$, denotes the successful transmission of the selected message at two instants of time, $i$ and $i + j$. This event can only occur if no other users transmit during the time interval ${-K + i, -K + i + 1, \dots, i + j}$.
Therefore, if both $F_i$ and $F_{i+j}$ occur, then all intermediate events $F_l$ for $i < l < i + j$ must also occur; that is, the message must have been successfully transmitted at all intermediate time instants as well.
Therefore
\begin{align}
\label{thu}
P(F_i F_{i+j}) &=\P(F_iF_{i+1} \dots  F_{i+j}) 
\nonumber
\\
&=\frac{1}{(1+q)^{N-1}} (1-q)^{(N-1)(K+j)}(1-q)^{i+j-1},    
\end{align}
regardless of whether there are events $F_l,$  $i< l < i+j$ in the intersection or not.

Let us count the number of distinct events corresponding to successful transmissions between two fixed time instants.
We begin with a single event of the form $F_iF_{i+j}$, indicating successful transmissions at times $i$ and $i+j$.
Next, for $j>1$, there are $\binom{j-1}{1}$  events of the form $F_iF_{m}F_{i+j},$, where $i<m<i+j$.
For $j>2$ we can further consider $\binom{j-1}{2}$ events of the form $F_iF_{m}F_{n}F_{i+j},$ where $i<m<n<i+j$, etc.

Therefore, the event $F_iF_{i+j}$ for $j\leq K-i$ will be included in the inclusion-exclusion formula $(\ref{incl})$ with coefficient

\begin{align*}
-(1-\varepsilon)^2 &+ \binom{j-1}{1} (1-\varepsilon)^3  - \binom{j-1}{2} (1-\varepsilon)^4 + \dots 
\\
&+ (-1)^{j}(1-\varepsilon)^{j+1} = -(1-\varepsilon)^2 \varepsilon^{j-1}. 
\end{align*}
    
Then formula $(\ref{incl})$ can be rewritten in the following form:  
\begin{align*}
& \P\left(\bigcup_{i=0}^K F_iD_i\right) = (1-\varepsilon)\sum_{i=0}^K \P(F_i)
 \\
 &- (1-\varepsilon)^2\sum_{j=1}^K \varepsilon^{j-1}\sum_{0\leq i\leq K-j} \P(F_iF_{i+j}) 
 \\
&=(1-\varepsilon) \frac{(1-q)^{(N-1)K}}{(1+q)^{N-1}} 
\\
&\times{}
\left(1+ \frac{1-(1-q)^K}{q}
- (1-\varepsilon)
\frac{(1-q)^{N-1}}{q} \right.
\\
& \left.
\times \left( \frac{1- \left(\varepsilon (1-q)^N\right)^{K}}{1-\varepsilon (1-q)^{N}} 
- (1-q)^K \frac{1- \left(\varepsilon (1-q)^{N-1}\right)^{K}}{1-\varepsilon (1-q)^{N-1}}\right)\right).
\end{align*}

By direct transformations, this expression can be reduced to the form presented in the statement of Theorem~\ref{T1}, which completes the proof.

\section{Appendix 2: Comments to the proof of Theorem  \ref{T2}}

Consider an individual message that appears at time $n = 0$, and analyze the probability of its successful delivery to the BS. 

As before, the message is considered to be delivered to the BS if it is successfully transmitted and not corrupted by random noise during transmission. Therefore, the probability of delivery can be represented as  
\[
\P\left(\bigcup_{i=0}^K F_iD_i\right).
\]

The same reasoning as in the proof of Theorem~$\ref{T1}$ applies. As in the previous case, the occurrence of the event $F_i F_{i+j}$, where $1 \leq j \leq K - i$, implies that the selected message is successfully transmitted at times $i$ and $i + j$. This event can occur only if no other messages are present at the time instants $-K + i, -K + i + 1, \dots, i + j$, 
\[
F_i F_{i+j} =\{\xi_{-K+i} = 0,\xi_{-K+i+1} = 0,\dots, \xi_{i+j} = 0\}.
\]
Hence, if both events $F_i$ and $F_{i+j}$ occur, then necessarily all intermediate events $F_l$ for $i < l < i + j$ must also occur. Therefore, in this case, formula~(\ref{thu}) takes the following form:
\begin{align*}
\P(F_i F_{i+j}) =\P(F_i \dots F_{i+j}) = e^{-\lambda (K+j-i+1)}    
\end{align*}
regardless of whether the intermediate events
$F_l$, for $i < l < i + j$, explicitly appear in the intersection. Consequently, formula~(5) reduces to the expression stated in Theorem~2.


\end{document}